\documentclass[fleqn,usenatbib]{mnras}

\usepackage{newtxtext,newtxmath}

\usepackage[T1]{fontenc}

\DeclareRobustCommand{\VAN}[3]{#2}
\let\VANthebibliography\thebibliography
\def\thebibliography{\DeclareRobustCommand{\VAN}[3]{##3}\VANthebibliography}

\usepackage{graphicx}	
\usepackage{amsmath}	

\newcommand{\jzjc}{$j_z/j_{\mathrm{circ}}$}
\newcommand{\meanjzjc}{$\langle j_z/j_{\mathrm{circ}}\rangle$}
\newcommand{\jpjc}{$j_{\rm{p}}/j_{\mathrm{circ}}$}

\newcommand{\emin}{$e/e_{\mathrm{min}}$}
\newcommand{\kappaco}{$\kappa_{\mathrm{co}}$}

\newcommand{\eagle}{{\sc{Eagle}}}
\newcommand{\subfind}{{\sc{Subfind}}}

\newcommand{\fdisk}{$f_{\rm{disc}}$}
\newcommand{\fihl}{$f_{\rm{SH}}$}
\newcommand{\fexsitu}{$f_{\rm{ex\ situ}}$}
\newcommand{\mus}{$\mu_{{\rm peak}}$}
\newcommand{\exs}{{\em ex situ}}
\newcommand{\ins}{{\em in situ}}

\title[The impact of mergers on stellar haloes]{The weak connection between the stellar haloes and merger histories of Milky Way-mass galaxies}

\author[Proctor et al.]{
Katy L. Proctor,$^{1,2,3}$\thanks{E-mail: katy.proctor@fysik.su.se} Aaron D. Ludlow$^{1}$, Claudia del P. Lagos$^{1,2}$, Aaron S. G. Robotham$^{1,2}$
\\
$^{1}$International Centre for Radio Astronomy Research (ICRAR), M468, University of Western Australia, 35 Stirling Highway, Crawley, WA 6009, Australia. \\
$^{2}$ARC Centre of Excellence for All Sky Astrophysics in 3 Dimensions (ASTRO 3D).\\
$^{3}$The Oskar Klein Centre, Department of Physics, Stockholm University, 106 91 Stockholm, Sweden.}

\date{Accepted XXX. Received YYY; in original form ZZZ}

\pubyear{2025}

\begin{document}
\label{firstpage}
\pagerange{\pageref{firstpage}--\pageref{lastpage}}
\maketitle



\begin{abstract}
Stellar haloes form through the disruption of satellite galaxies over time, making them a promising observable for constraining galaxy merger histories.
We use a dynamical decomposition technique to isolate the stellar haloes of Milky Way-mass galaxies in the $100\,{\rm Mpc}$ \eagle\ simulation and study their relationship to the merger histories of their hosts.
We define the stellar halo as the stellar mass that is bound to the central subhalo but not associated with the disc or bulge components of a galaxy, and we quantify their merger histories using the most significant merger since $z=1$.
Surprisingly, we find that the fraction of a galaxy's total stellar mass in the stellar halo, $f_{\rm SH}$, is not a reliable indicator of its merger activity.
Contrary to common assumptions, disc galaxies with low $f_{\rm SH}$ do not necessarily have quiescent merger histories.
In fact, roughly one quarter experienced a merger at $z \leq 1$ with a satellite whose stellar mass was at least 10 per cent of the host galaxy's stellar mass.
These galaxies undergo mergers with satellites on circular orbits that are roughly co-planar with the pre-existing disc and thereby avoid contributing mass to the stellar halo.
Instead, such mergers build thick, extended discs and supply fresh gas that often triggers a significant episode of star formation in the disc.
Our results suggest that disc galaxies with low-mass stellar haloes, such as the Milky Way, can have varied and active merger histories, and that stellar haloes may not be a reliable tool for inferring galaxy merger histories.
\end{abstract}


\section{Introduction}
A defining prediction of the Lambda Cold Dark Matter ($\Lambda$CDM) model is that galaxies form in part through mergers with satellite galaxies.
Mergers can contribute significantly to the stellar component of galaxies, either through the direct provision of stars and cold gas \citep[e.g.][]{de_lucia_hierarchical_2007} or by igniting rapid bursts of star formation in merger remnants \citep[e.g.][]{barnes_shock-induced_2004, springel_formation_2005}. 
Mergers are the dominant mechanism by which massive galaxies assemble their stellar mass \citep[e.g.][]{robotham_galaxy_2014} and are often invoked as the primary means by which disc galaxies are transformed into elliptical galaxies \citep[e.g.][]{toomre_mergers_1977, hopkins_mergers_2010, martin_role_2018}. 
Understanding the extent to which galaxy merger histories can be inferred from present-day observables is an active area of research in astrophysics \citep[e.g.][]{eisert_ergo-ml_2022, rey_how_2022}.
 
Stellar haloes may offer powerful constraints on galaxy merger histories due to their natural connection to hierarchical assembly.
As the most extended stellar component of galaxies and dynamically-distinct from centrally-concentrated discs and bulges, stellar haloes are thought to form primarily from the debris of disrupted satellites \citep[e.g.][]{bullock_tracing_2005, cooper_galactic_2010}. 
Observationally, stellar haloes are challenging to characterise due to their low surface brightness, particularly in external galaxies. 
Nonetheless, existing observations reveal considerable galaxy-to-galaxy variation in stellar halo properties. 
For example, the stellar halo mass fraction ($f_{\rm SH}$) among Milky Way–mass disc galaxies spans nearly an order of magnitude \citep[e.g.][]{merritt_dragonfly_2016}, a variation often attributed to stochasticity in their merger histories. 
Galaxies with low $f_{\rm SH}$ are typically assumed to have quiescent merger histories, while those with high values are assumed to have more active pasts \citep[e.g.][]{elias_stellar_2018, smercina_saga_2020, gozman_m94_2023}.
For example, M31 has a massive stellar halo that exhibits prominent substructure consistent with a major merger occurring as recently as $z \approx 0.15$ \citep{dsouza_infall_2021}.

The Milky Way’s stellar halo has been studied in far greater detail than any other, owing to the availability of detailed chemo-dynamical data. 
The Gaia-Sausage-Enceladus (GSE) is a prominent substructure in the Milky Way’s stellar halo thought to be the remnant of a massive satellite galaxy accreted around 10 Gyr ago \citep{belokurov_co-formation_2018, helmi_merger_2018}.
Observational signatures such as the break in the stellar halo density profile are also consistent with the GSE resulting from a single major accretion event \citep[][]{deason_apocenter_2018}, reinforcing the picture of the Milky Way as a relatively quiescent galaxy that has not undergone significant merger activity since the GSE event. 
However, this interpretation has been challenged by recent studies of the chemodynamics of inner halo stars that suggest the MW may have experienced an additional, more recent merger with a relatively massive dwarf galaxy \citep[e.g.][]{donlon_ii_milky_2020, donlon_debris_2024, horta_stellar_2024, liu_gse_2025}.

The number and timing of mergers that contributed to the MW remain uncertain, highlighting the challenge of interpreting the properties of stellar haloes as direct indicators of merger history.
Establishing the origin of observed variations in stellar halo properties requires input from simulations, where causal inferences can be drawn between the formation of stellar haloes and past merger events.
Cosmological simulations now reproduce the observed properties of galaxies reasonably well \citep[e.g.][]{schaye_eagle_2015, pillepich_first_2018}, yet two outstanding technical issues have hampered efforts to build a comprehensive picture of stellar halo formation in a $\Lambda$CDM universe.

First, stellar haloes are typically the least massive and thus most poorly-resolved component of a galaxy.
In MW-mass galaxies, stellar haloes typically constitute $\lesssim 10$ per cent of the total stellar mass \citep[e.g.][]{elias_stellar_2018, canas_stellar_2020, proctor_identifying_2024}.
As a result, theoretical studies of stellar haloes often make use of "zoom" simulations in which the mass resolution can be increased while maintaining cosmologically-realistic formation histories \citep[e.g.][]{pillepich_building_2015, bonaca_gaia_2017, sanderson_reconciling_2018, yu_stars_2020, rey_how_2022}. 
The high resolution of such simulations comes at a cost: a limited number of galaxies can be studied, and they may inherit selection biases \citep[for example, candidates for zoom simulations are often selected based on environmental conditions, or using strict isolation criteria, e.g.][]{grand_auriga_2017}.

The second issue pertains to characterising the stellar halo.
A variety of methods are employed in theoretical studies, and
even those making use of the 6-D phase space information provided by simulations struggle to unambiguously distinguish stellar halo particles from those belonging to a centrally concentrated disc or bulge \citep[see e.g.][]{canas_stellar_2020}.
Other studies opt to first define a disc component using a cut in the orbital circularity of stellar particles, and then distinguish the bulge from the stellar halo using a threshold in binding energy \citep[e.g.][]{zolotov_dual_2009, mccarthy_global_2011}.
While these approaches may be useful for small samples of galaxies for which the appropriate thresholds can be determined by eye, \cite{liang_connection_2024} showed that the optimal threshold in orbital circularity (binding energy) used to define the disc (stellar halo) component of galaxies in cosmological simulations exhibits substantial galaxy-to-galaxy variation, even for galaxies of the same stellar mass.
Accurately identifying the stellar haloes (and discs and bulges) of galaxies that form in large volume cosmological simulations requires a more nuanced approach.

The primary aim of our study is to investigate the link between the stellar halo components of MW-mass galaxies and their assembly histories. We address the aforementioned technical issues by applying a dynamical decomposition technique \citep[first described in][]{proctor_identifying_2024} to identify the structural components of galaxies in the flagship $(100\, {\rm Mpc})^3$ \eagle\ simulation \citep{schaye_eagle_2015}.
Our decomposition technique was designed to isolate the stellar haloes of galaxies that span a broad range of mass and morphology, and the large volume of the \eagle~simulation offers a broad range of galaxy formation histories. This allows us to investigate the extent to which the stellar halo encodes recent accretion events.

The remainder of the paper is organized as follows.
In Section \ref{sec:methods}, we present the simulations and sample of galaxies used in our study, as well as the analysis techniques used to quantify galaxy merger histories, morphologies, and stellar components. 
In Section \ref{sec:results}, we quantify the impact of peak mergers on galaxy morphologies, stellar halo fractions, and the present-day \ins\ and \exs\ stellar mass content of galaxies. In Section \ref{sec:conc} we summarize our results.
\section{Simulations and Analysis}\label{sec:methods}

\subsection{Simulations}
Our analysis focuses on galaxies identified in the 
$(100 \,{\rm Mpc})^3$ intermediate resolution \eagle~simulation \cite[see][for details]{schaye_eagle_2015,crain_eagle_2015}. 
The initial conditions were sampled using ${\rm N}=1504^3$ dark matter and gas particles 
that were evolved to $z=0$ using a modified version of the {\sc gadget3} code \citep{springel_cosmological_2005}.
The cosmological parameters are consistent with those determined by the 
\cite{planck_collaboration_planck_2014}. 
The dark matter and gas particle masses are therefore $m_{\rm dm}=9.70\times 10^6\,{\rm M}_\odot$ and
$m_{\rm g}=1.81\times 10^6\,{\rm M}_\odot$, respectively; the (Plummer-equivalent) gravitational
softening length is $\epsilon=0.7\,{\rm kpc}$ in physical units (at $z\leq 2.8$; $\epsilon=2.66$ comoving kpc
at higher redshifts).

Due to its finite resolution, \eagle~employs subgrid physics modules for among other processes,
radiative cooling and photoheating \citep{wiersma_chemical_2009}, the growth of black holes through mergers and accretion \citep{rosas-guevara_impact_2015}, star formation and mass loss due to stellar evolution \citep{schaye_relation_2008}, and
energetic feedback from stars and active galactic nuclei \cite[see][for details]{schaye_eagle_2015}. 
The subgrid model parameters were calibrated so that \eagle~reproduced a set of $z=0$ observations of the galaxy population, including their stellar mass function and size-mass relation, although it also reproduces a variety of observations not used for calibration \citep[see][for details]{crain_eagle_2015}. 
Our analysis uses the simulation carried out with the ``Reference'' subgrid model parameters.

\subsection{Halo finding and merger trees}\label{subsec:merger_histories}

DM haloes and their associated galaxies were identified using {\sc subfind} \cite[][]{springel_populating_2001, dolag_substructures_2009}.
For each halo, \subfind~identifies the position of the particle with the minimum potential
energy, which we identify as its centre. 
We define the virial mass, ${\rm M_{200}}$, as the total mass
within a sphere surrounding the halo's centre that encloses a density contrast equal to
$\rho_{200}\equiv 200\times \rho_{\rm crit}$. The virial radius is thus defined by
$r_{200}=(3\,{\rm M}_{200}/[4\, \pi\, \rho_{200}])^{1/3}$. 
The total stellar mass of each galaxy, ${\rm M_\star}$,
is the mass of all stellar particles that \subfind~deems bound to the central DM halo but not to
any of its satellite galaxies. Likewise, ${\rm M}_{\rm g}$ is the mass of gas particles bound to each galaxy.
We have verified our results are qualitatively unchanged if we limit our definition to the total mass within a fixed physical aperture, for example $30\,{\rm kpc}$.

We connect galaxies to their progenitors at earlier epochs using the merger trees described in
\cite{mcalpine_eagle_2016} and \cite{qu_chronicle_2017}. 
At a given snapshot, each galaxy has a unique main progenitor in the preceding snapshot,
which is defined as the one with the largest `branch mass' \citep[i.e., the galaxy with largest total mass summed across all previous snapshots, see][]{de_lucia_hierarchical_2007}.
All other progenitors are
classified as satellite galaxies, some of which may eventually merge with the main progenitor. When
necessary, we denote quantities pertaining to satellite galaxies using ``sat'' subscripts;
we do not use subscripts for quantities pertaining to central galaxies or their main progenitors. 

We classify the stellar particles of each central galaxy as \ins~ if they are
bound to its main progenitor in the first snapshot after their formation or as \exs\ if they formed in satellites of the main progenitor or in an external galaxy. The fraction of
\exs\ (\ins) stellar mass is denoted $f_{\rm ex\ situ}$ ($f_{\rm in\ situ}$).

\subsection{The galaxy sample}

Our study focuses on 1254 central galaxies with $z=0$ virial masses in the range
$10^{11.8}\, {\rm M}_{\odot} \leq {\rm M}_{200} \leq 10^{12.3}\, {\rm M}_{\odot}$.
This approximately covers the range of MW masses reported in recent literature \citep[see fig. 6 of][]{shen_mass_2022},
but also ensures that their relevant properties are robust to the effects of spurious collisional
heating, which can alter the structure and morphology of simulated galaxies
\citep[see][for details]{ludlow_spurious_2021, wilkinson_impact_2023, ludlow_spurious_2023}. 
\cite{proctor_identifying_2024} demonstrated that the disc, bulge, and stellar halo masses for \eagle~galaxies
in this mass range are unaffected by spurious collisional effects.
Note that we impose no relaxation or
isolation criteria on our sample.

\subsection{Classifying the present day morphology and structural components of \eagle~galaxies}

We adopt a coordinate system centred on each galaxy (or its main progenitor) and align the $z-$axis with the total angular momentum
vector of its stellar component. Thus, $j_{z,i}$ is the $z-$component of the angular
momentum vector of particle $i$, $R_i=(x_i^2+y_i^2)^{1/2}$ is its distance from the $z-$axis, and
$r_i=(R_i^2+z_i^2)^{1/2}$ is its 3D distance from the galaxy centre.

We decompose each galaxy into three possible structural components: a disc, a bulge,
and a stellar halo, by applying a 12-component Gaussian mixture model (GMM) to
the circularities (\jzjc\ and \jpjc) and specific binding energies (\emin) of their stellar particles.
Specifically, we search for clusters of stellar particles in this parameter space and approximate them by
multi-dimensional Gaussian distributions. Each particle has a non-zero probability of being associated with
all of the 12 Gaussians, but we assign it wholly to the Gaussian component to which it has the
highest assignment probability (this is sometimes referred to as a ``hard'' assignment scheme). We then associate
each Gaussian distribution to a physical galaxy component as follows.
Those with \meanjzjc$\geq 0.5$ (note: \meanjzjc~is the mean value of each Gaussian along the \jzjc~axis)
are assigned to the disc component; the remainder are assigned to either the bulge or stellar halo depending
on whether their mean specific binding energy, $\langle e/e_{\rm min} \rangle$, is above (for the bulge) or below (for the stellar halo)
$e_{\rm cut}=[\max(\langle e/e_{\rm min}\rangle)+\min(\langle e/e_{\rm min} \rangle)]/2$. 
Note that under our definition, all components (discs, bulges and stellar haloes) can contain \ins\ and \exs\ stellar particles.
In this work, we use the subscripts ``disc'', ``bulge'' and ``SH'' to distinguish quantities pertaining
to different galactic components. For example, $f_{\rm disc}$ is the fraction of mass in the disc relative to the total
stellar mass of the galaxy; \fihl\ is the stellar halo mass fraction, and so on.
More information
about our dynamical decomposition method can be found in \cite{proctor_identifying_2024}.

We quantify the $z=0$ morphology of each galaxy using the disc mass fraction,
i.e. \fdisk\, $={\rm M_{disc}}\, / \,{\rm M_\star}$ (the fraction of stellar
mass assigned to the bulge or stellar halo will be denoted $f_{\rm bulge}$ or $f_{\rm SH}$, respectively).
\citet[][see their fig. 5]{proctor_identifying_2024} showed that
\fdisk\ is strongly correlated with the kinematic morphology indicator \kappaco,\footnote{Specifically,
$\kappa_{\rm co}=(2\,K_\star)^{-1}\sum_{j_{z,i}>0} m_i\,(j_{z,i}/R_i)^2$, where $K_\star$ is the total
kinetic energy of the stellar particles.} defined as the fraction of stellar kinetic energy
in co-rotation with the galaxy. Because $\kappa_{\rm co}$ is easier to calculate than \fdisk, we will use it
to characterise galaxy morphologies at $z>0$. 

We identify $z=0$ disc galaxies as those with \fdisk$\,\geq 0.4$ (roughly equivalent to the convention used by
\citealt{correa_dependence_2020} and identical to the MW-like selection applied in \citealt{aschersleben_gamma_2024}) but note that our results are insensitive to reasonable variations in this
threshold; adopting \fdisk$\geq 0.5$, for example, yields similar results.

\subsection{Quantifying recent mergers}

Our goal is to study the effect of recent mergers on the stellar haloes of MW mass
galaxies.
We use the merger trees described in Section \ref{subsec:merger_histories} to identify the most massive merger that each central
galaxy has undergone since $z=1$.\footnote{We consider mergers that occur at $z\leq 1$ to preclude the epoch
during which the progenitors of most MW-mass galaxies are only a small fraction of their present-day mass
and are frequently undergoing mergers. Our results are qualitatively unchanged if we adopt a slight higher
merger redshift cutoff, $z\leq 2$ for example.}
To do so, we first identify all secondary progenitors that have completely merged\footnote{We assume that a
satellite galaxy has completely merged when it is no longer detected by \subfind.} with each main progenitor 
and from them identify the one that had the maximum stellar mass at any point during its history;
we will refer to this as the ``peak'' merger and denote the progenitor's mass as ${\rm M_{peak,sat}}$
and the redshift at which occurs as $z_{\rm peak}$.
Of the 1254 haloes in our sample, 23 do not experience any merger events in the selected time frame. These haloes are included in our analysis, but excluded from figures presenting the properties of merging satellites.

The stellar mass ratio of the peak merger is defined as \mus\,$\equiv {\rm M_{peak,sat}}/{\rm M}_\star(z_{\rm peak})$,
where ${\rm M}_\star(z_{\rm peak})$ is the stellar mass of the main progenitor at $z_{\rm peak}$.
\citep[see e.g.][for a similar definition of merger mass ratios]{rodriguez-gomez_merger_2015, qu_chronicle_2017}.

For a small number of galaxies ($\approx 5$ per cent) we find the above procedure results in \mus $\gg 1$
as a result of subhalo swapping events that sometimes occur when two galaxies of comparable mass are near
coalescence or peri-centric passage when a snapshot is written \citep[see e.g.][]{poole_convergence_2017}. 
We redefine these spurious \mus\ values based on the time when the merging galaxy attained its {\em second} highest
stellar mass, which usually occurs in the preceding snapshot. If the corrected \mus\ value is less
than the original value (and not $\gg 1$) we use it instead to quantify the peak merger event.
We have visually inspected all mergers for which this correction was applied and found it yields sensible
\mus\ values for all but 5 galaxies in our sample. 
These galaxies have anomalous merger trees (due to numerical issues)
and were excluded from further analysis.

Various other properties of the peak merger event that we use throughout this paper include: 1) the infall redshift,
$z_{{\rm infall}}$, at which the merging galaxy was first identified as a satellite of the main progenitor; 
2) the redshift $z_{{\rm merge}}$ at which the merging satellite was last identified as a distinct galaxy by \subfind;
3) the alignment angle, $\cos\theta$, between the main progenitor's stellar angular momentum vector and the
orbital angular momentum vector of the merging satellite (both measured at $z_{{\rm merge}}$); 
4) the ratio of pericentre ($r_{{\rm p}}$) to apocentre ($r_{{\rm a}}$) of the merging satellite\footnote{We estimate $r_{{\rm p}}$ and $r_{{\rm a}}$ by obtaining the roots of the radial energy equation assuming an NFW potential with the same values of $V_{{\rm max}}$ and $R_{{\rm max}}$ as that of the central galaxy's halo \citep[see eq. 3.14 of][]{binney_galactic_2008}.} measured at the snapshot prior to $z_{{\rm merge}}$; 
and 5) the gas fraction,
$f_{{\rm g, sat}}=M_{{\rm g, sat}}/ (M_{{\star, \rm sat}} + M_{{\rm g, sat}})$, of the merging satellite at $z_{{\rm infall}}$.

Note that our definitions of $z_{{\rm infall}}$ and $z_{{\rm merge}}$ depend on the time cadence of the \eagle\ snapshots, which vary from $\approx 0.3-1$ Gyr. In this work, we are primarily interested in determining whether a galaxy experienced a significant merger at $z<1$, and so we do not anticipate the limited time cadence of the simulation output will affect our results (as defined above, our derived values of $z_{{\rm merge}}$ represent upper limits on the true merger time).
\section{Results}\label{sec:results}

\subsection{The imprint of merger history on the morphologies of Milky Way mass galaxies}

\begin{figure}
  \centering
  \includegraphics{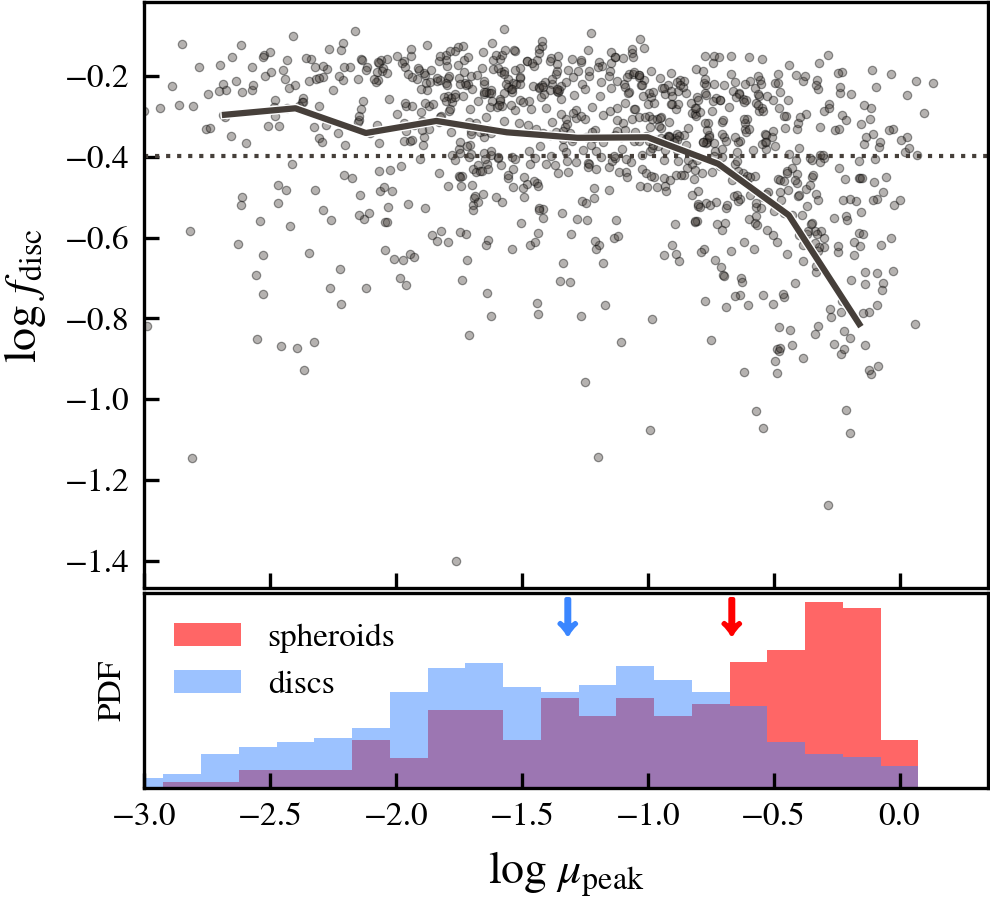}
  \caption{Disc mass fraction, $f_{\rm disc}$, plotted as a function of $\mu_{\rm peak}$, the stellar mass ratio of the peak merger event since $z=1$
    (top panel). Points correspond to individual galaxies and the solid line shows the median trend.
    The horizontal dashed line at \protect{\fdisk $=0.4$} corresponds to the lower limit for disc galaxies. 
    The bottom panel shows the distribution of $\mu_{\rm peak}$ for spheroids (in red; i.e. those with \protect{\fdisk $=0$})
    and discs (in blue; i.e. those with \protect{\fdisk $\geq 0.4$}). The downward pointing arrows of 
    corresponding colour denote median $\mu_{\rm peak}$ values for these populations.}
  \label{fig:morph_mu}
\end{figure}

The top panel of Fig. \ref{fig:morph_mu} plots the relation between \fdisk\ and \mus; 
the solid line is the median relation. Galaxies with quiet merger histories ($\mu_{\rm peak}\lesssim 0.1$) are typically
(though not always) disc dominated, whereas those with active merger histories ($\mu_{\rm peak} \gtrsim 0.1$)
have a large spread of disc mass fractions. This supports the view that minor and major mergers
(i.e. \mus$\gtrsim 0.1$)
are generally destructive to discs, while mini mergers
(i.e. \mus$\lesssim 0.1$)
typically have little impact on disc morphology \citep[e.g.][]{martin_role_2018, trayford_star_2019} and have even been shown to contribute to the growth of stellar discs in \eagle\ galaxies \citep[e.g.][]{clauwens_three_2018}. 
Mini mergers are typically poorly-resolved: for \mus$\lesssim 0.1$, most satellites are resolved with fewer than 120 stellar particles\footnote{These satellites are, however, typically resolved with over 3000 DM particles, and the stellar and halo mass functions of \eagle\ galaxies are converged on these scales \cite[see][for further details]{ludlow_numerical_2020}.}. Including these low-mass mergers in our analysis does not affect our conclusions, which are primarily based mergers with  \mus$\gtrsim 0.1$, and correspond to progenitors resolved by 1000s of stellar particles (and even more dark matter particles).

It is clear from the bottom panel of Fig. \ref{fig:morph_mu} that both discs and spheroids exhibit a broad
range of merger histories. Here we plot the distributions of \mus\ values for galaxies with \fdisk$=0$ (red histogram; spheroids, hereafter)
and for those with \fdisk$\geq 0.4$ (blue histogram; discs, hereafter). While spheroids typically have more active merger histories
than discs (the median \mus\ values are 0.045 and 0.198 for discs and spheroids, respectively),
many discs have experienced significant recent mergers while many spheroids have not. For
example, $\approx 34$ per cent of discs have $\mu_{\rm peak} \geq 0.1$ and $\approx 38$ percent of spheroids have $\mu_{\rm peak} \leq 0.1$.

The diverse merger histories exhibited by galaxies of similar virial mass and morphology has been reported before.
\cite{sales_origin_2012} showed that major mergers are not a prerequisite for spheroid formation -- misaligned gas
accretion can instead lead to episodic star formation resulting in stellar populations with misaligned angular momenta
and little coherent rotation. 
\cite{garrison-kimmel_origin_2018} showed that disc galaxies can host ex situ stellar components whose orbits are, in some cases, closely aligned with those of the in situ disc.
Additionally, \citet[][see also, e.g., \citealt{lagos_angular_2017, sotillo-ramos_merger_2022}]{font_diversity_2017}
showed that stellar discs can form or regrow after major mergers, provided the remnant retains a substantial gas supply and has
a relatively quiescent merger history thereafter. 

The morphologies of MW-mass galaxies therefore provide an inadequate proxy for their
recent merger histories, which leads us to search for clues elsewhere. Given their claimed sensitivity to hierarchical galaxy assembly \citep[e.g.][]{bullock_tracing_2005, monachesi_auriga_2019},
such clues may lie in the stellar haloes of galaxies. We turn our attention to this next.

\subsection{The stellar haloes, \exs~masses, and merger histories of Milky Way mass galaxies}

\begin{figure}
  \centering
  \includegraphics{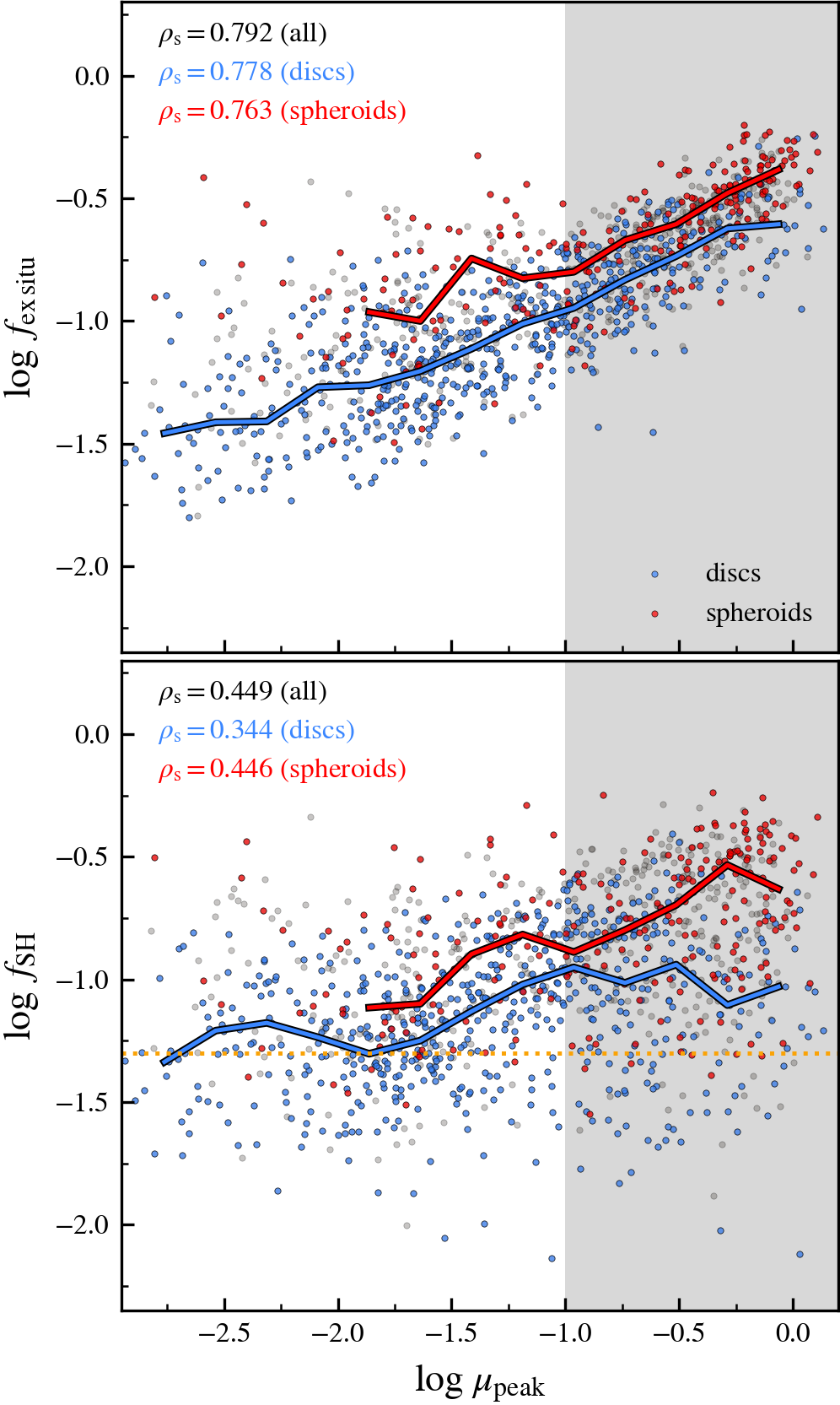}
  \caption{The upper (lower) panel plots the relation between \protect{\fexsitu} (\protect{\fihl}) and \protect{\mus}.
    Blue points correspond to disc galaxies ($f_{\rm disc}\geq 0.4$), red points to spheroids ($f_{\rm disc}=0$), and
    grey points to all other galaxies; the median
    relations for discs and spheroids are shown using lines of corresponding colour. The Spearman rank correlation coefficients, \protect{$\rho_s$}, for
    each relation are given in the top left corner of each panel.}
  \label{fig:corrs}
\end{figure}

The top panel of Fig. \ref{fig:corrs} shows the relationship between \fexsitu~and \mus. Blue and red
points distinguish discs and spheroids, respectively, and grey points show the remaining galaxies. The blue and red 
lines show the median relations for discs and spheroids, respectively. 
The Spearman rank correlation coefficients ($\rho_s$, labelled in the upper left corner) show
that these two quantities are strongly correlated for both morphologies ($\rho_s=0.76$ for spheroids and $\rho_s=0.78$ for discs),
but at fixed \mus, spheroids have slightly higher \exs~fractions than discs. This is not only because discs typically
experience lower-mass peak mergers (Fig.~\ref{fig:morph_mu}) but also because they experience fewer mergers overall.
In addition, discs typically continue to form stars \ins~after their peak mergers, whereas star formation is usually
depleted in spheroids \citep[e.g.][]{bellstedt_galaxy_2020}. 

As expected,
galaxies that have experienced a significant recent merger have higher \exs~fractions than those that have not. In fact,
the peak mergers described above are typically the dominant contributors to the present day \exs~stellar mass of most
galaxies, at least at the MW mass scale. This agrees with the conclusions of \cite{fattahi_tale_2020}, who showed that the majority of
the accreted stellar mass found in the central regions of galaxies can often be traced back to a single merger.

{\em Ex situ} mass cannot be directly observed but stellar haloes can be (at least in principle), and are assumed to be
dominated by \exs~stars. Nevertheless, regardless of morphology,
the relation between $\mu_{\rm peak}$ and $f_{\rm SH}$ plotted in the bottom panel of Fig.~\ref{fig:corrs} is significantly weaker
than the $f_{\rm ex-situ}-\mu_{\rm peak}$ relation, suggesting only a fraction of the accreted material ends up in a loosely bound stellar halo.

Another interesting difference between the two relations is that systems with low-mass stellar haloes (e.g. $f_{\rm SH}\lesssim 0.05$, those below the horizontal dashed line in the bottom panel) -- which are typically disc galaxies -- 
span the entire range of \mus~values plotted, with a few having $\mu_{\rm peak}\gtrsim 0.3$.
While it is clear from
the top panel of Fig. \ref{fig:corrs} that having a low fraction of \exs~stellar mass essentially precludes recent mergers of substantial mass, the
bottom panel shows that having low \fihl\ does not. 
In fact, roughly a quarter of discs with $f_{\rm SH}\leq 0.05$ undergo peak
mergers 
with \mus$\, \geq 0.1$.

This challenges the conventional
view that galaxies with discy morphologies and low-mass stellar haloes have experienced quiet merger histories
\citep[e.g.][]{elias_stellar_2018, jang_tracing_2020, du_physical_2024, ma_evolutionary_2024}, suggesting that stellar haloes have
limited power to constrain galaxy assembly histories. Together, the results plotted in the top and bottom panels of
Fig.~\ref{fig:corrs} suggest
that a significant amount of accreted material can instead make its way into the disc or bulge components of galaxies, and that
under some conditions, significant mergers may go by unnoticed by the stellar halo.

Table \ref{tab:subsamples} provides some information about the galaxy samples considered in our paper, including
the number of discs and spheroids, the number of discs with $f_{\rm SH}\leq 0.05$, and how many of these have
active ($\mu_{\rm peak} > 0.1$) or quiescent merger histories ($\mu_{\rm peak} \leq 0.1$).

\begin{table}
    \centering
    \caption{Number of galaxies separated by $z=0$ morphology and merger history.}
    \begin{tabular}{l | rrr}
    \hline
                                &                            &    Morphology              &            \\
    \hline      
    Merger class &  discs (\fihl\ $<0.05$) &  discs &  spheroids \\
    \hline
    active ($\mu_{\rm peak} \geq 0.1$) &                         43 &              210 &        153 \\
    quiescent ($\mu_{\rm peak} < 0.1$) &                        145 &              405 &        91 \\
    \hline
    \end{tabular}
    \label{tab:subsamples}
\end{table}

\subsection{The origin of disc galaxies with low-mass stellar haloes and active merger histories}\label{subsec:merger}

\begin{figure*}
  \centering
  \includegraphics{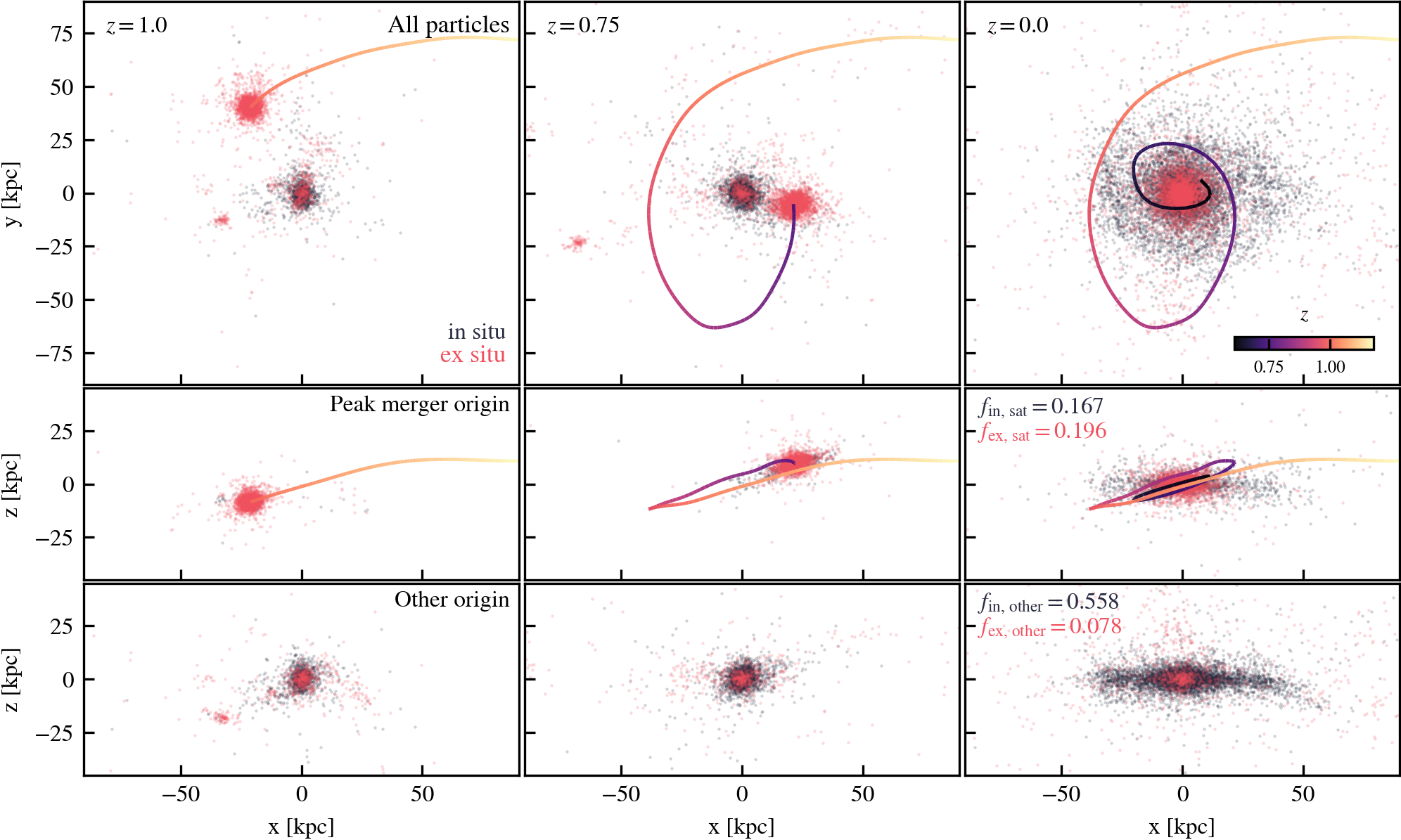}
  \caption{Projections of stellar particles during a major merger. The left and middle columns
    show the distributions $\approx 2\,{\rm Gyr}$ and $\approx 0.4\,{\rm Gyr}$ before $z_{\rm merge}\approx 0.6$,
    respectively; the rightmost column shows the present-day distribution. The top row shows face-on projections
    of all stellar particles that formed prior to these times, provided they are bound to the central galaxy at $z=0$.
    Black and red points are the subsets of \ins~and \exs~stellar particles, respectively. The middle row shows
    edge-on projections of the subset of stellar particles associated with the merging satellite: red points are \exs~
    star particles born in the satellite and later stripped off; black points are star particles born \ins~from
    stripped gas particles originally belonging to the satellite.
    The trajectory of the satellite is displayed in the top and middle rows, colour-coded by redshift.
    The bottom row shows edge-on projections of all
    other \ins~(black) and \exs~(red) stellar particles. The orientation of the merger with the growing disc is
    such that the majority of \exs~stellar mass from the peak merger ends up in the disc and bulge (32 per cent and 67 per cent, respectively) of the descendant galaxy 
    rather than in the stellar halo.}
  \label{fig:projs}
\end{figure*}


How do disc galaxies with low-mass stellar haloes arise from the diverse merger histories implied by Fig. \ref{fig:corrs},
in particular those succumbing to major mergers? In Fig.~\ref{fig:projs} we consider an example of such a galaxy. Here we
plot the spatial distribution of stellar particles at different phases of a major merger,\footnote{The merger contributes
$\approx 36$ per cent of the present-day stellar mass of the galaxy, with roughly equal contributions from
\exs~star particles stripped from the satellite ($\approx 72$ per cent of the galaxy's total \exs~mass), and
from stripped gas particles that later form stars \ins~($\approx 23$ percent of the galaxy's total \ins~mass).
The peak merger mass ratio is $\mu_{\rm peak}=0.91$.} whose descendant
has a massive disc component ($f_{\rm disc}=0.464$) and a relatively low-mass stellar halo ($f_{\rm SH}=0.044$). The leftmost and
middle columns show the particle distributions $\approx 1.9$ and $\approx 0.4\,{\rm Gyr}$ before $z_{\rm merge}$, respectively, and the rightmost
column shows the present day stellar mass distribution.
The top row shows the face-on projections, whereas the middle
and bottom rows show edge-on projections.
In all panels, black and red points correspond to \ins~ and \exs~star particles, respectively.
The edge-on projections are further split into stellar particles that were directly contributed
by the peak merger (middle row), and those that were not (bottom row). More specifically, the red points in the middle row
correspond to stellar particles that were stripped from the satellite; black points correspond to stellar particles born \ins~from
gas particles that were stripped from the satellite. All other \ins~and \exs~star particles are shown in the
bottom row.

The satellite approaches the main progenitor on a low inclination -- though misaligned -- orbit ($\cos \theta = -0.58$ at $z=1$) that becomes
more aligned with the angular momentum vector of the central galaxy as the merger progresses ($\cos\theta = 0.98$ by $z_{\rm merge}$).
By $z=0$ the stellar particles contributed by the peak merger (red points, rightmost middle panel) occupy a flattened distribution
that is co-planar with the \ins\ disc. Contrast this with the remaining \exs~stellar particles (bottom-right panel),
which are more centrally concentrated and spherically distributed.
The \ins~star particles are aligned with the disc regardless of whether or not they originate from the merging satellite
(as seen in the middle- and lower-right hand panels).

The co-planar configuration of mergers such as this are an important factor
contributing to the survival of massive discs \citep[see, e.g.][]{zeng_formation_2021} and the formation of thin discs \citep[see, e.g.][]{hu_formation_2024}. 
The example plotted in Fig.~\ref{fig:projs}
suggests that the remnants of such mergers may also have relatively low-mass stellar haloes, since the majority of \exs~
material inherits the orbital angular momentum of the merging satellite, and therefore occupies a rotationally-supported
disc rather than contributing to a dispersion supported spheroid reminiscent of a classical stellar halo.

\begin{figure}
    \centering
    \includegraphics{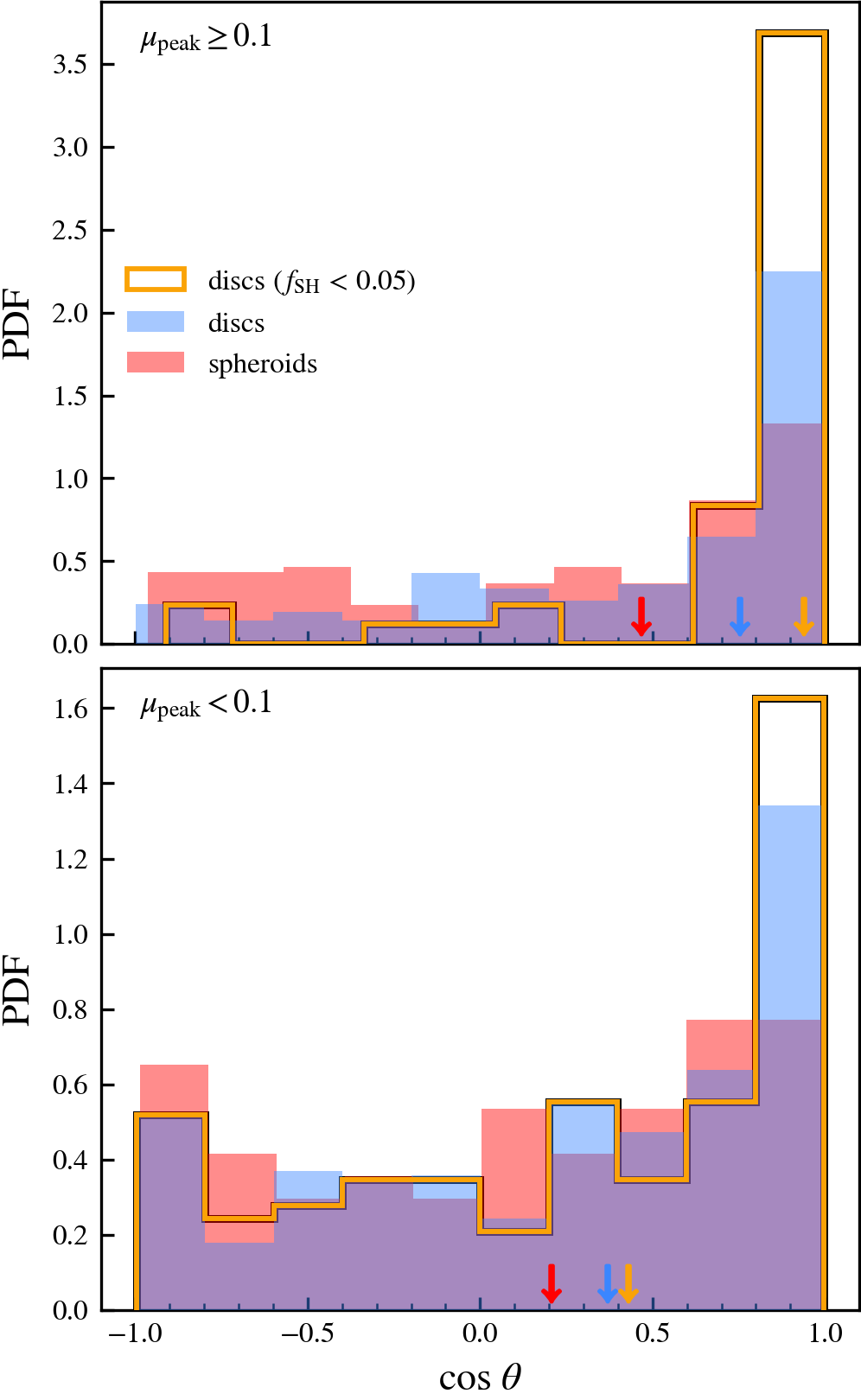}
    \caption{Distributions of $\cos\theta$, the angle between the orbital angular momentum vector of the peak merger
    and the total stellar angular momentum vector of the central, both measured at \protect{$z_{{\rm merge}}$}. The top and bottom
    panels show results for peak mergers with mass ratios \mus$\geq 0.1$ and \mus$<0.1$, respectively. The filled
    red and blue histograms correspond to spheroids and discs, respectively; the open orange histograms correspond
    to discs with $f_{\rm SH}\leq 0.05$. The medians of each distribution are indicated by arrows or corresponding colour.}
  \label{fig:alignment}
\end{figure}

\begin{figure}
    \centering
    \includegraphics{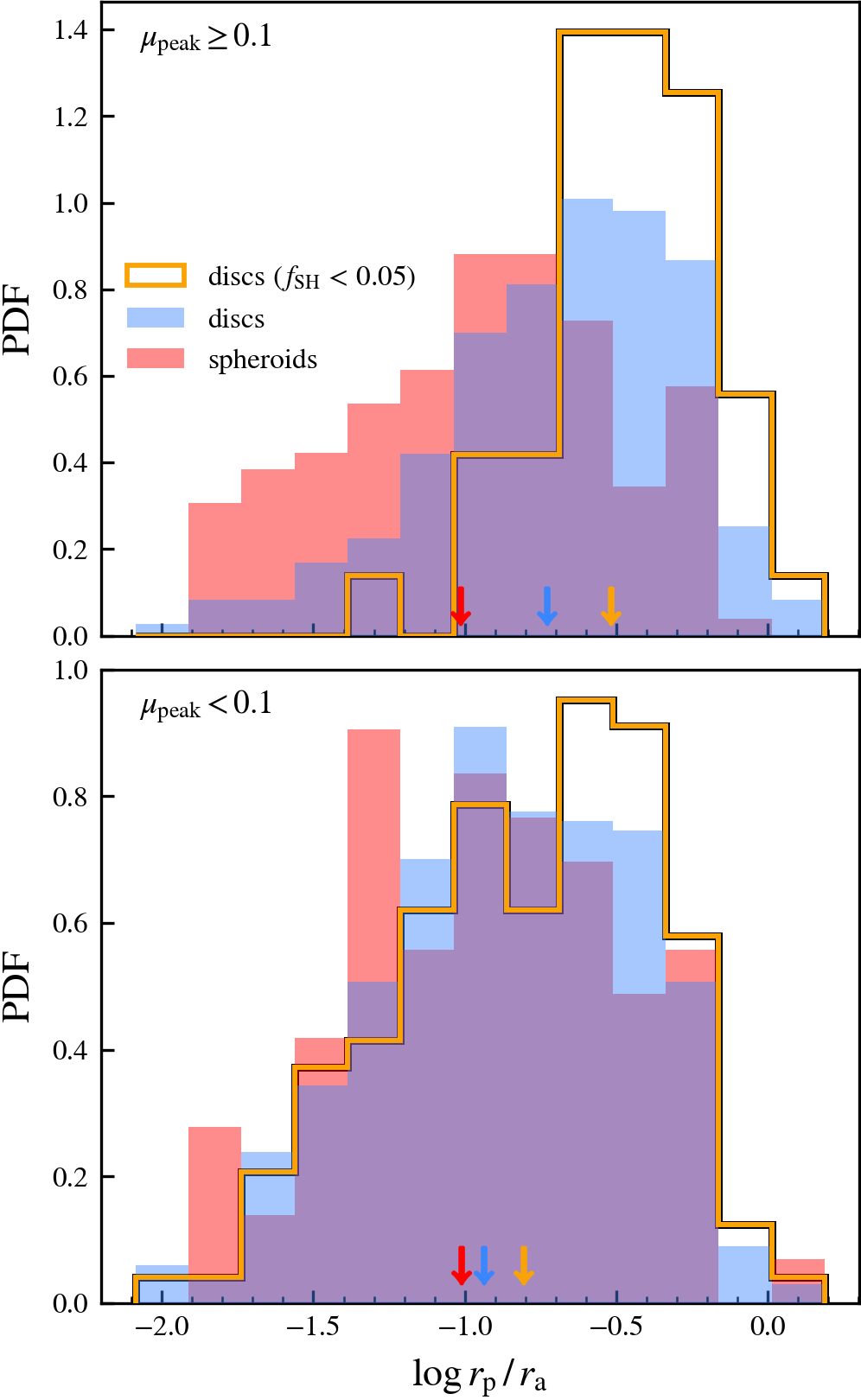}
    \caption{Distributions of $r_{\rm p}/r_{\rm a}$, the ratio of predicted pericentre and apocentre of the peak merger satellite, measured at the snapshot immediately preceding \protect{$z_{{\rm merge}}$}. The top and bottom
    panels show results for peak mergers with mass ratios \mus$\geq 0.1$ and \mus$<0.1$, respectively. The filled
    red and blue histograms correspond to spheroids and discs, respectively; the open orange histograms correspond
    to discs with $f_{\rm SH}\leq 0.05$. The medians of each distribution are indicated by arrows or corresponding colour.}
  \label{fig:eccs}
\end{figure}

In Fig. \ref{fig:alignment} we plot the distribution of $\cos \theta$ (measured at $z_{\rm merge}$) for all peak merger events 
experienced by galaxies with active (\mus $\geq 0.1$; top panel) and quiescent (\mus$<0.1$; bottom panel) merger histories.
Results for discs and spheroids are plotted using blue and red histograms, respectively, and the subset of discs with
$f_{\rm SH}\leq 0.05$ are plotted in orange.
Note that isotropic accretion of satellites onto a spherical surface would result in a uniform $\cos\theta$ distribution.

Peak mergers for disc galaxies have a strong tendency to be aligned with the disc plane. The alignment is strongest for $\mu_{\rm peak}\geq 0.1$
(on average, $\langle\cos\theta\rangle = 0.75$) but is also noticeable for the lower-mass mergers
($\langle\cos\theta\rangle=0.36$ for $\mu_{\rm peak}<0.1$). 
When limited to discs with $f_{\rm SH}\leq 0.05$ the alignment
probability becomes even stronger, with $\langle\cos\theta\rangle=0.92$ for $\mu_{\rm peak}\geq 0.1$\footnote{The few
$\mu_{\rm peak}\geq 0.1$ mergers that are not co-planar in the $f_{\rm SH}\leq 0.05$ sample typically occur at early
times, when the main progenitors have not yet built up a significant stellar disc component. In these cases,
  the peak merger contributes little to the total $z=0$ stellar mass of the galaxy.} 
and $\langle\cos\theta\rangle=0.36$ for $\mu_{\rm peak}< 0.1$. 
Indeed, for galaxies experiencing $\mu_{\rm peak}\geq 0.1$ mergers, a Kolmogorov–Smirnov (KS) test suggest that the $\cos\theta$ distribution for the peak mergers of discs with $f_{\rm SH}\leq 0.05$ is significantly different from the distributions of both the full disc population and spheroids, with KS statistics of 0.31 and 0.47, respectively (both tests return p-values $<0.005$).
  
Although spheroids also show a preference for aligned mergers, the effect is much weaker than for discs.
The alignment is particularly weak for spheroids experiencing peak mergers with $\mu_{\rm peak}< 0.1$, where the $\cos \theta$ distribution is statistically consistent with isotropic accretion (i.e. a KS test under the null hypothesis that the $\cos \theta$ distribution is identical to a uniform distribution returns a p-value of 0.09).

Fig. \ref{fig:eccs} plots the ratio of the pericentre and apocentre of the peak merger orbits at the snapshot immediately prior to $z_{\rm merge}$. For \mus $> 0.1$, peak mergers for disc galaxies also tend to have more circular orbits than spheroids, with $\langle r_{\rm p}/r_{\rm a}\rangle=0.1$ and $\langle r_{\rm p}/r_{\rm a}\rangle=0.18$, respectively.
The peak mergers for discs with low stellar halo masses (i.e. $f_{\rm SH}\leq 0.05$) are the most circular of all, with $\langle r_{\rm p}/r_{\rm a} \rangle=0.30$.

The results shown in Figs.~\ref{fig:alignment} and \ref{fig:eccs} suggest that the orbital trajectory of a merging satellite plays a significant role in
determining the location of material in the descendant.
Numerous studies based on controlled simulations of individual mergers have shown that satellites on orbits
that are approximately co-planar with the disc of their host galaxy can result in the formation of an extended, thick disc
component home to a significant fraction of accreted stars
\citep[e.g.][]{abadi_simulations_2003, penarrubia_formation_2006}. 
Fig. \ref{fig:alignment} suggests that this
scenario is not only plausible, but is common in a cosmological setting, particularly for MW-mass disc galaxies with low-mass
stellar haloes. 

\begin{figure}
  \centering
  \includegraphics{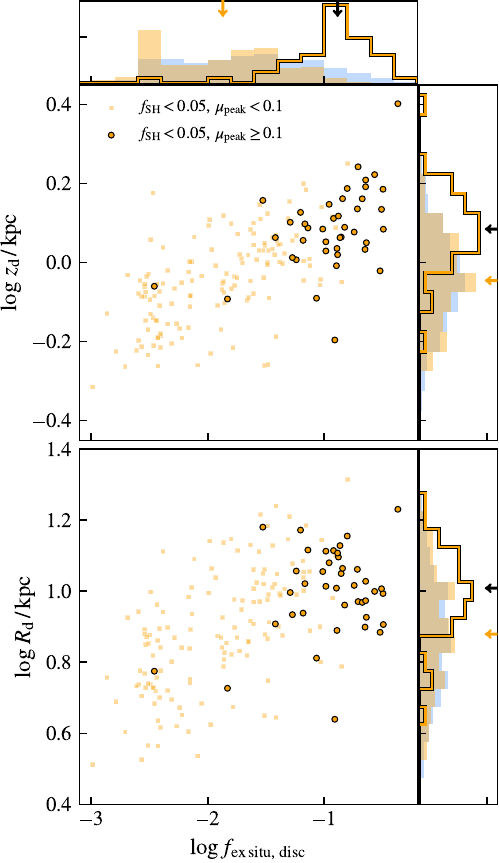}
  \caption{The upper (lower) panel plots the vertical half-mass scale height (length) of the disc component as a function of the \exs~mass fraction of the disc component
    for disc galaxies with \fihl$\,<0.05$. 
    Galaxies with \fihl$\,<0.05$ and quiescent histories are plotted with transparent yellow squares; galaxies  with \fihl$\,<0.05$ and active histories are plotted with filled circles. The distribution of \exs~fractions and scale heights/lengths for galaxies with \fihl$\,<0.05$ are shown in yellow in the upper
    and right panels, respectively. Black and yellow arrows denote the median values for the active and quiescent sample, respectively. The distributions for the full disc galaxy population is shown for reference and is plotted in blue}
  \label{fig:disc_heights}
\end{figure}

To ascertain the effect that co-planar mergers may have on the structure of the disc component, in Fig. \ref{fig:disc_heights} we plot the half-mass scale heights and lengths of discs with \fihl$<0.05$, as a function of the disc's {\em ex situ} mass fraction (i.e. the fraction of the disc mass that formed {\em ex situ}).
The small pale points correspond to galaxies with $\mu_{\rm peak}<0.1$ whereas larger points correspond to those with $\mu_{\rm peak}\geq 0.1$ (i.e., those that have experienced recent co-planar mergers).
As expected, galaxies with active merger histories have thicker and more extended discs and higher \exs~fractions, indicating that more massive 
mergers result in the formation of thicker discs. 

In addition to being larger in size than the quiescent disc sample, we note that the disc component of galaxies experiencing mergers with $\mu_{\rm peak}\geq 0.1$ are typically younger than their quiescent counterparts, with median half-mass stellar ages of  5.3 and 6.4 Gyrs, respectively. 
These results are in qualitative agreement with disc properties inferred from spectral decomposition of GAMA galaxies by \cite{robotham_profuse_2022}.
These authors show that, at fixed stellar mass, the largest discs typically host the youngest stellar populations.
Our results provide a potential interpretation of this result: disc galaxies with active merger histories tend to have larger and younger discs than quiescent galaxies, indicating that co-planar mergers may also rejuvenate star formation in the disc by supplying fresh gas.
This suggests that, at fixed stellar mass, the most extended discs are a result of recent co-planar mergers that incite star formation in the remnant galaxy. 
We explore the total contribution of the peak merger to the \ins\ and \exs\ stellar mass content of the central galaxy further in the next section.

\subsection{Peak mergers and their contributions to {\em in situ} and {\em ex situ} stellar mass}

\begin{figure}
  \centering
  \includegraphics{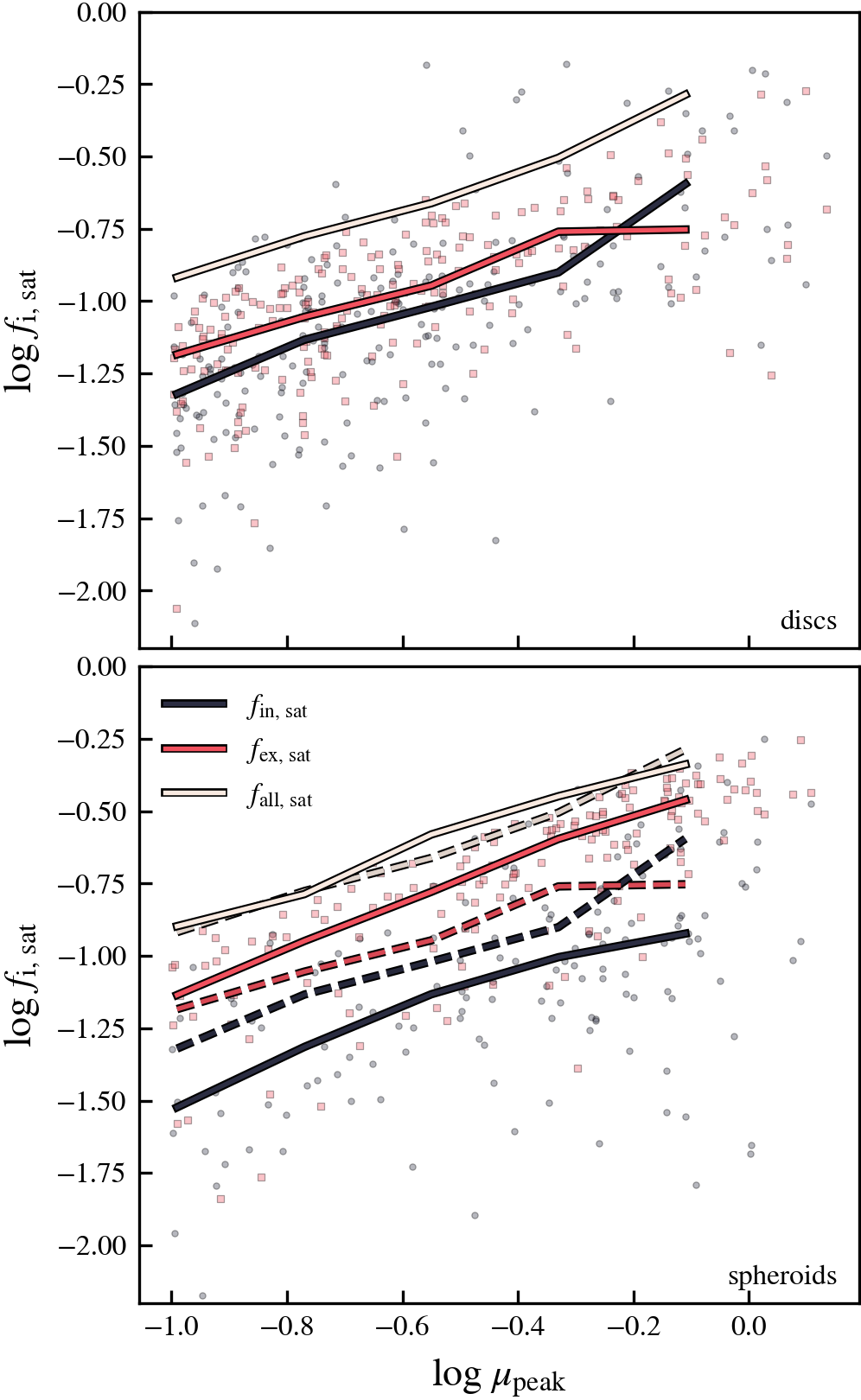}
  \caption{The fraction of stellar mass in central galaxies contributed by their peak mergers and plotted as a
    function \mus. The contribution from \exs~star particles are shown using pink squares; \ins~star
    particles (i.e. those formed from gas particles brought in by the merger) are shown using black circles.
    The median relations are plotted as lines of corresponding colour, with the total median contribution plotted
    in white. The top and bottom panels show the relations separately for discs and spheroids, respectively.
    For comparison, we also plot the median relations for disc galaxies in the bottom panel using dashed lines.}
  \label{fig:contr}
\end{figure}

The example shown in Fig.~\ref{fig:projs} suggests that major mergers not only contribute a substantial fraction to
a galaxy's total \exs~mass, but can also supply gas reserves that incite \ins~star formation after the merger has completed.
In Fig.~\ref{fig:contr} we plot the fractional contributions of peak mergers to the total mass of their descendants as a
function of $\mu_{\rm peak}$ (for $\mu_{\rm peak}\geq 0.1$). The top and bottom panels correspond to $z=0$ discs and
spheroids, respectively. 
The fractional contributions are divided between \exs~star particles
contributed directly by the merger ($f_{\rm ex,sat}$; shown in red) and \ins~star particles triggered by gas that the merger supplied ($f_{\rm in,sat}$; shown in black). 
Note that in both cases the
masses are relative to the total $z=0$ stellar mass of the central galaxy.
The red and black lines show the
median trends for \exs~and \ins~mass, respectively; the white line shows the median trend for the total mass
fraction contributed by the merger ($f_{\rm tot,sat}$).
The relative contribution of the peak merger increases with \mus, with higher mass ratio mergers contributing more to
the present day \exs~and \ins~stellar mass.

The top panel of Fig.~\ref{fig:contr} shows that peak mergers typically contribute an
equal amount of \exs~and \ins~stellar mass to disc galaxies. 
The bottom panel shows that peak mergers have a different effect on spheroids
(the dashed lines are the median relations for discs, repeated in this panel for comparison). In this case, the peak merger
contributes more significantly by the direct provision of stripped stellar particles than by stripped gas particles that later
form stars \ins. Overall, spheroids have higher fractions of \exs~stellar mass than discs and lower fractions of 
\ins~stellar mass.

Whether a merging satellite contributes more to the \exs~or \ins~stellar mass of a galaxy will be
determined by both the baryonic properties of the satellite at infall and its subsequent evolution within the halo of the main
progenitor. For example, gas poor mergers (${\rm M}_\star \gtrsim {\rm M}_{\rm g}$ at infall) will likely contribute more to
the \exs~mass of the galaxy than to its \ins~mass, since most of its baryons are in the form of stellar particles before
the merger ensues. Conversely, gas rich mergers (${\rm M_g}\gtrsim {\rm M}_\star$ at infall) can in principle contribute
significantly to the \ins~stellar mass of the galaxy by replenishing the central's fuel supply.

For gas rich mergers, however, the situation can be quite complicated. For example, the merging
satellite may proceed to form stars prior to completely merging with the host galaxy, and if its star formation rate
is sufficiently high, its contribution to the \exs~stellar mass budget may outweigh that of the \ins~mass regardless of
its gas fraction at infall (recall that star particles formed within the satellite are considered \exs~by our definition).
On the other hand, given sufficient time, stripped gas particles can cool and be converted to star particles within the
main progenitor, leading to an
enhanced contribution to its \ins~stellar mass. Which of these two scenarios dominates is likely a non-trivial function of
the satellite's orbit within its host, its gas fraction and star formation rate at infall, and whether feedback from the
host is able to prevent stripped gas particles from cooling and forming star particles \ins.

The tendency displayed in Fig. \ref{fig:contr} for galaxies to experience a roughly equal, or greater contribution of \exs~stars
suggests that, at least for spheroids, the peak merger satellites either remain star-forming throughout their orbits, or that the gas stripped
from the satellite is prevented from cooling and forming stars through internal baryonic processes such as feedback.

\begin{figure}
  \centering
  \includegraphics{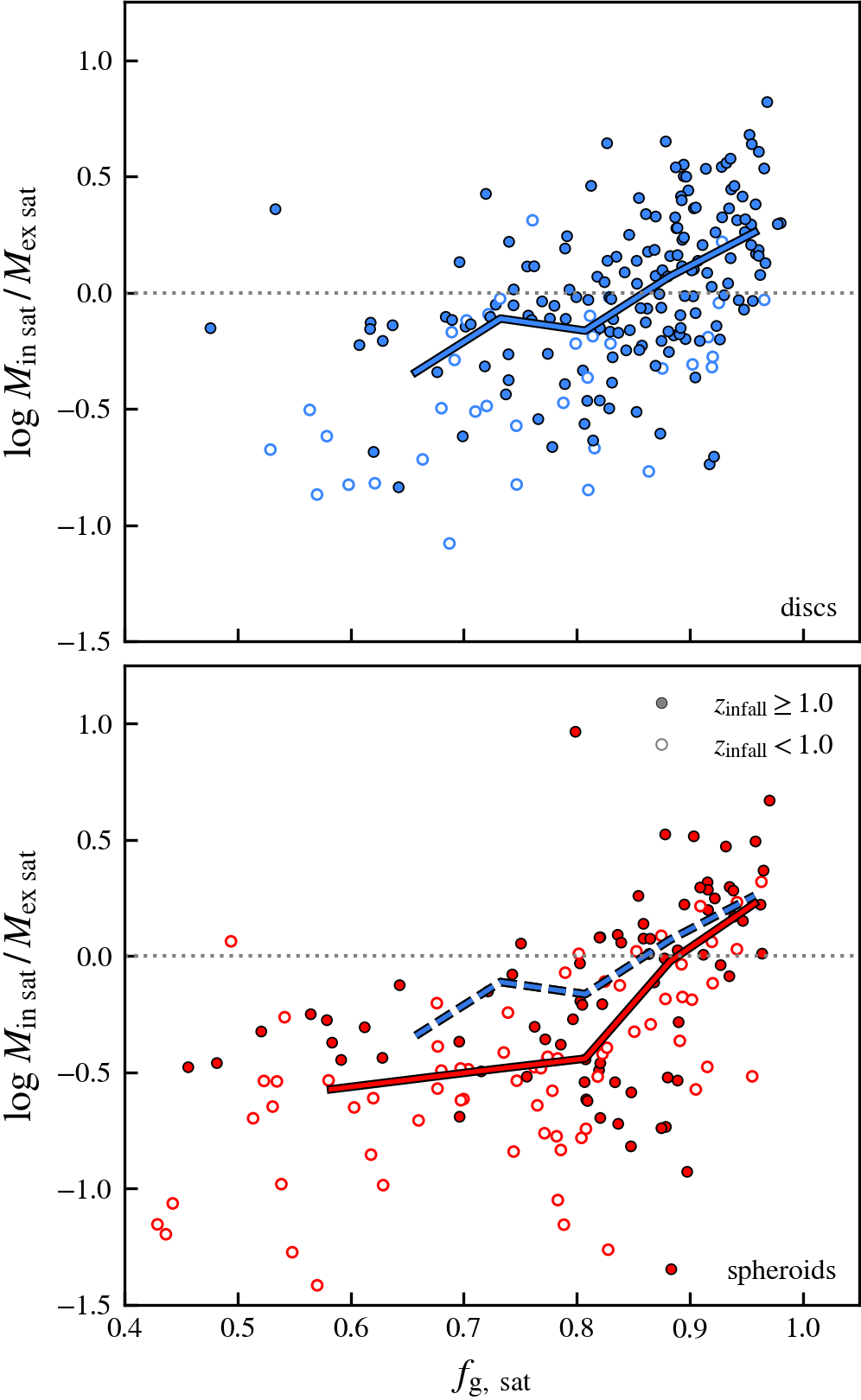}
  \caption{Ratio of the \ins~and \exs~masses contributed by the peak merger to our sample of discs (top panel) and
    spheroids (bottom panel). Results are plotted as a function of the satellite's gas fraction, $f_{\rm g, sat}$, at
    infall and are colour-coded by the infall redshift (filled circles for $z_{\rm infall}\geq 1.$ and empty circles
    for those with $z_{\rm infall}<1$; $z_{\rm infall}=1$ is the median infall redshift for the discs and
    spheroids in our sample). Note that ${\rm M_{ex,sat}}$ is the fraction of central's stellar mass contributed
    directly by the peak merger in the form of \exs~star particles; ${\rm M_{in,sat}}$ is the fraction of the
    central's stellar mass that formed \ins~from gas particles stripped from the satellite.}
  \label{fig:pme_contr_inex}
\end{figure}

In Fig. \ref{fig:pme_contr_inex} we examine the relative contribution of the peak merger to the \exs~and \ins~stellar mass of
the central galaxy, and how it depends on the satellite's gas fraction at infall.
Specifically, we plot the ratio of the \ins~stellar mass triggered by the merger (${\rm M_{in,sat}}$) to the \exs~stellar mass stripped
from the satellite (${\rm M_{ex,sat}}$) as a function of $f_{\rm g,sat}$. 
Points correspond to individual galaxies and are coded by infall time (open circles for satellites with $z_{\rm infall}<1$;
filled circles for those with $z_{\rm infall}\geq 1$; note that $z_{\rm infall} = 1$ corresponds to
the median infall redshift of all $\mu_{\rm peak} \geq 0.1$ peak mergers.). As in Fig.~\ref{fig:contr}, the relation is plotted
separately for discs (blue points; upper panel) and spheroids (red points; lower panel).

As expected, the relative contribution of the peak merger to the \exs~and \ins~ is correlated with $f_{{\rm g,sat}}$:
satellites with higher gas fractions typically contribute more \ins~mass than \exs~mass.
Note too the relationship between $f_{\rm g,sat}$ and $z_{\rm infall}$: satellites that are accreted earlier do so with higher
gas fractions. This ensures that there is sufficient time for early accreted, gas rich satellites to be stripped of their
gas that can later go on to form \ins~stars in the central galaxy. This interpretation is supported by the segregation of points by
infall time that can be readily seen in both panels of Fig.~\ref{fig:pme_contr_inex}: gas poor satellites with late infall times
typically have $M_{\rm ex,sat} > M_{\rm in,sat}$, whereas gas rich satellites with early infall times have
$M_{\rm ex,sat} < M_{\rm in,sat}$. This is consistent with the findings of \cite{lagos_angular_2017}, who showed that the gas fraction
of mergers typically decreases with decreasing redshift. 

\section{Discussion and Conclusions}\label{sec:conc}

We used the $(100\, {\rm Mpc})^3$ simulation from the \eagle\ project \citep{schaye_eagle_2015} to study the connection between the merger histories and stellar haloes of Milky Way-mass galaxies ($10^{11.8}\, {\rm M}_{\odot} \leq {\rm M}_{200} \leq 10^{12.3}\, {\rm M}_{\odot}$). 
Galaxies were separated from their stellar haloes using the structural decomposition method developed by \cite{proctor_identifying_2024}, which assigns stellar particles to galactic components (discs, bulges, or stellar haloes) based on their dynamics.
This approach allowed us to distinguish between kinematically distinct components, such as a thick, extended disc and a stellar halo, that can be difficult to separate using traditional decomposition techniques \citep[see e.g.][]{liang_connection_2024}. 

Applying this method within a large-volume cosmological simulation also allowed us to analyse galaxies that formed in diverse environments and experienced a wide range of merger histories. We defined the peak merger as the most massive merger since z = 1, quantified by the stellar mass ratio (\mus). 
This served as a proxy for merger activity, allowing us to distinguish between galaxies with quiescent (\mus < 0.1) and active (\mus > 0.1) merger histories.

Our main result is that a high stellar halo mass fraction (\fihl) reliably indicates an active merger history, independent of galaxy morphology. However, low \fihl\ values do not necessarily imply a quiescent history, limiting the reliability of stellar haloes as indicators of past mergers. 
For example, many disc galaxies with low-mass stellar haloes have experienced significant mergers: approximately 25 per cent of discs with \fihl \,$<0.05$ underwent a merger with $\mu_{\rm peak}>0.1$ within the past 8 Gyr. 
These mergers tend to occur in configurations that are coplanar with a pre-existing disc, and deposit substantial amounts of ex situ mass into the disc while leaving the stellar halo largely unaffected.

This suggests that the Milky Way's low stellar halo mass cannot be used to exclude a major merger in its recent past; instead, merger remnants may be embedded in its disc. 
These results are relevant to recent studies proposing that the Milky Way’s last major merger involved a direct collision with the Galactic disc, as inferred from chemo-dynamical signatures in the local stellar halo \citep[see e.g.][]{donlon_pendulum_2023, donlon_debris_2024, liu_gse_2025}.

Our analysis shows that co-planar mergers can occur without significantly disrupting present-day disc morphology or stellar halo content.
The primary structural difference between disc galaxies with and without active merger histories lies in the disc itself: galaxies with \mus\,$>0.1$ tend to have thicker and more radially extended discs than those with $\mu_{\rm peak}<0.1$. 
While further study is needed to determine whether these structural differences are observable, upcoming deep surveys---such as the Wide-Area VISTA Extragalactic Survey \citep{driver_4most_2019} and the European Space Agency’s Euclid Wide Survey \citep{scaramella_euclid_2021, euclid_collaboration_euclid_2022}---combined with spectral decomposition techniques \citep[e.g.][]{robotham_profuse_2022} may offer valuable observational constraints.

Although stellar haloes are more difficult to characterise observationally, recent work suggests that morphological features within them \citep[e.g. tails, shells and streams][]{khalid_characterizing_2024, valenzuela_stream_2024} or the shapes of stellar density profiles \citep[e.g.][]{pillepich_halo_2014, deason_apocenter_2018} may provide additional insight into past merger events.
We also stress that our results were obtained using a dynamical decomposition scheme, which cannot be compared directly to observations that rely on photometric light profile fitting, which often suffer from uncertainty due to the potential confusion between stellar haloes and thick disc components \citep[see e.g.][for a discussion]{merritt_dragonfly_2016}.

Nevertheless, our findings underscore a key point: low mass stellar haloes do not necessarily imply a quiescent merger history, challenging a common assumption in the literature \citep[e.g.][]{elias_stellar_2018, smercina_saga_2020, gozman_m94_2023, ma_evolutionary_2024}.

Our main results can be summarized as follows:

\begin{enumerate}

\item The peak merger mass ratio \mus\ is strongly correlated with the present day \exs~mass fraction for all
  galaxies, regardless of their morphology. This suggests that the most massive merger occurring since $z \leq 1$
  is the dominant contributor to the \exs~mass of most MW-mass galaxies and that, for galaxies that experience peak mergers with \mus\, $\geq 0.1$, lower-mass or
  earlier mergers do not contribute much to \fexsitu.

\item Stellar haloes are typically believed to be dominated by \exs~mass and yet their stellar mass fractions, \fihl,
  correlate more weakly with \mus~than does the total \exs~mass fraction (Fig. \ref{fig:corrs}). Additionally, we showed
  that galaxies with low-mass stellar haloes are not necessarily those with quiescent merger histories.
  For example, roughly one quarter of disc galaxies with \fihl$<0.05$ have experienced a merger with \mus$\, > 0.1$ since $z=1$. This suggests that galaxies with low-mass stellar haloes and disc-dominated morphologies do not necessarily have quiet merger histories, as is commonly assumed; indeed, many do not.

\item The vast majority of disc galaxies with low-mass stellar haloes and active merger histories undergo peak mergers with satellites whose orbits are more circular than average and whose orbital angular momentum is aligned with the stellar angular momentum vector of the central galaxy at $z_{\rm merge}$ (Fig. \ref{fig:alignment} and Fig. \ref{fig:eccs}). Co-planar mergers with \mus$>0.1$ do not contribute much \exs~mass to the stellar halo, but instead deposit their mass into the disk and bulge components (see Fig. \ref{fig:disc_heights}). Present-day discs that undergo such mergers are generally thicker, more extended, and have higher fractions of \exs~material than those with quiescent merger histories. 

\item Peak mergers with $\mu_{{\rm peak}}\gtrsim 0.1$ contribute a significant fraction to the $z=0$ stellar mass of a galaxy
  (Fig. \ref{fig:contr}), as much as $\approx 40$ per cent for the most massive mergers. Part of this mass is in
  the form of \exs~stellar material formed within the satellite and brought into the central galaxy during the merger;
  the remainder is in the form of \ins~stellar material originating from gas particles stripped from the satellite during the merger.
  For disc galaxies (and $\mu_{{\rm peak}}\geq 0.1$), the \exs~and \ins~stellar mass contributed by the peak merger are approximately equal.
  For spheroids, the contribution of \exs~stellar mass usually exceeds that of the \ins~stellar mass, except among the most gas rich mergers.

\item Differences in the relative amounts of \exs~and \ins~contributed by peak mergers can be attributed to differences in
  the gas-richness of the satellite at infall (see Fig. \ref{fig:pme_contr_inex}). 
  Gas-rich peak mergers -- more common among our sample of disc galaxies -- typically contribute more \ins~than \exs~stellar
  mass, and generally occur at higher redshifts than gas poor mergers, which are more common among the spheroid population. 
  As a result, peak mergers for disc galaxies generally contribute more \ins~than \exs~stellar material to the central
  galaxy, whereas for spheroids the opposite is true. 
  
\end{enumerate}

Finally, we have shown that ($z\leq1$) peak mergers often account for a significant fraction of the present-day mass of discs without contributing significantly to their stellar haloes, but have paid little attention to the properties of the progenitors that do contribute to the stellar halo component.
A comprehensive analysis of the progenitors of the stellar halo, disc, and bulge components of \eagle\ galaxies will be presented in a forthcoming paper.


\section*{Acknowledgements}

KLP acknowledges support from the Australian Government Research Training Program Scholarship.
ADL acknowledges financial support from the Australian Research Council through their Future Fellowship scheme (project number FT160100250).
CL has received funding from the Australian Research Council Centre of Excellence for All Sky Astrophysics in 3 Dimensions (ASTRO 3D), through project number CE170100013, and the Australian Research Council Discovery Project (DP210101945).
ASGR acknowledges financial support from the Australian Research Council through their Future Fellowship scheme 
(project number FT200100375).

\section*{Data Availability}

The \eagle\ simulation data are publicly available; see \cite{mcalpine_eagle_2016} and \cite{the_eagle_team_eagle_2017} for further information. Additional data can be made available upon reasonable request.



\bibliographystyle{mnras}
\bibliography{paper} 




\appendix



\bsp	
\label{lastpage}
\end{document}